**Saggu, A., Ante, L., & Kopiec, K. (2024). Uncertain Regulations, Definite Impacts: The Impact of the US Securities and Exchange Commission's Regulatory Interventions on Crypto Assets. Finance Research Letters, 72, 106413.**

# Uncertain Regulations, Definite Impacts:
# The Impact of the U.S. Securities and Exchange Commission's Regulatory Interventions on Crypto Assets

Aman Saggu[ac]

[a] *Business Administration Division, Mahidol University International College, Mahidol University*

[c] *Blockchain Research Lab*

Lennart Ante[bc]

[b] *Constructor University*

[c] *Blockchain Research Lab*

Kaja Kopiec[c]

[c] *Blockchain Research Lab*

This Version: November 28, 2024

**Abstract:** This study employs an event study methodology to investigate the market impact of the U.S. Securities and Exchange Commission's (SEC) classification of crypto assets as securities. It explores how SEC interventions influence asset returns and trading volumes, focusing on explicitly named crypto assets. The empirical analysis highlights significant adverse market reactions, notably returns plummeting 12% over one week post-announcement, persisting for a month. We demonstrate that the severity of market reaction depends on sentiment and asset characteristics such as market size, age, volatility, and illiquidity. Further, we identify significant ex-ante trading volume effects indicative of pre-announcement informed trading.

**Keywords:** Cryptocurrency; Securities and Exchange Commission (SEC); Regulations; Event Study; Enforcement.

**JEL Classification:** G12; G14; G18; G15; K22.

# 1 Introduction

The U.S. Securities and Exchange Commission (SEC) significantly influences the cryptocurrency industry through unexpected regulatory interventions. SEC Chair Gary Gensler maintains a strict stance (SEC, 2023a), classifying most cryptocurrencies, except Bitcoin, as securities[1], claiming it protects investors and market integrity. However, this approach has sparked criticism, as SEC Commissioner Hester Pierce argues such stringent regulations stifle innovation and push entrepreneurs to more crypto-friendly jurisdictions (Pierce, 2024). At the heart of the debate lies the Howey Test, a legal standard from 1946 still used to determine whether instruments in the U.S. qualify as investment contracts and securities under U.S. law. The test's applicability to crypto assets' unique characteristics and complexity is contentious (Henning, 2018; Trotz, 2019). A key example is the case of major cryptocurrency Ripple (XRP). In 2020, the SEC filed charges against Ripple Labs, Inc. for selling unregistered securities, resulting in Ripple's delisting on the Coinbase cryptocurrency exchange.[2] In 2023, Ripple achieved a partial victory, upon which Coinbase relisted the asset (Coinbase, 2023). However, in October 2024, the SEC filed a notice of appeal to challenge this decision.

The lack of clear, comprehensive regulatory guidelines from the SEC for cryptocurrency projects has created an unpredictable, volatile environment fraught with risk and uncertainty for market participants. Technological advancements continue to spawn new asset types, complicating the SEC's mandate. However, the SEC's unexpected regulatory interventions (i.e., security classification, enforcement actions, advisory opinions, and comments) hamper the achievement of its mandate "to protect investors; maintain fair, orderly, and efficient markets; and facilitate capital formation" (SEC, 2023b), magnifying systemic risk and regulatory arbitrage. Contrary to public interest theory (Posner, 1974), the traditional U.S. financial industry may influence regulatory actions to limit the crypto sector to preserve market dominance (Stigler, 1974). This situation contrasts with the European Union, where the Markets in Crypto Assets Regulation (MiCa) provides greater clarity by categorizing crypto assets and setting specific regulatory standards (European Commission, 2020). Indeed, recent research reveals the SEC targets 'low-hanging fruit'—firms with conspicuous public trigger events and higher private sector scrutiny—to reduce their investigative burden (Holzman et al., 2024), a strategy contributing to unpredictable cryptocurrency regulatory interventions.

The existing literature reveals nuanced cryptocurrency market reactions to regulations. Auer and Claessens (2018) document strong negative market responses to Reuters news articles, between 2015-2018, about cryptocurrency treatment under securities law and positive reactions to new non-security cryptocurrency regulatory frameworks. In extreme cases of regulatory intervention, like the 2021 Chinese cryptocurrency ban, Griffith and Clancey-Shang (2023) identify a 41% market-wide crash persisting beyond 20 days post-announcement. However, Chen and Liu (2022) highlight the ban's ineffectiveness, finding that Chinese investors persisted in trading cryptocurrency despite the prohibition. Similarly, Borri and Shakhnov (2020) observed post-ban regional spillovers as trading volumes surged in Korea and Japan. Further exploring unintended consequences, Sauce (2022) argues such stringent international regulations push investors towards unregulated, non-compliant, decentralized venues beyond the regulatory radar.

---

[1] The SEC's approval of Ethereum ETFs in May 2025, categorized as "commodity-based trust shares," subtly suggests that Ethereum is considered more a commodity than a security.

[2] On July 13, 2023, the U.S. District Court for the Southern District of New York ruled that Ripple's trading on public exchanges did not violate any laws. However, Ripple's breached securities laws during sales of XRP to hedge funds and institutional investors.

Currently, event studies concerning cryptocurrency regulation principally leverage news coverage. Lyócsa et al. (2020) demonstrate that cryptocurrency regulation news in the Financial Times increased next-day volatility between 2013-2018, while Chokor and Alfieri (2021) identified negative abnormal returns between 2015-2019 following Factiva news coverage of future proposed cryptocurrency regulation. Similarly, between 2017 and 2019, Shanaev et al. (2020) observed a 1-3% decline in cryptocurrency portfolios following cryptocurrency regulation news globally. More recently, Bonaparte and Bernile (2023) constructed a Google Trends cryptocurrency regulation index, finding no significant long-term price impact on cryptocurrency returns but noted effects on volatility and volume.

Motivated by previous studies, this paper undertakes a unique investigation into the impact of SEC events classifying crypto assets as securities, focusing on the market impacts of explicitly named assets.[3] In pursuit of this objective, we formulate and empirically test two hypotheses: (H1) SEC classifications of crypto assets as securities are associated with negative abnormal returns following such regulatory announcements; (H2) Crypto assets' characteristics signaling market quality—such as larger market capitalization, older asset age, lower volatility, and higher liquidity—and positive external sentiment mitigate the negative abnormal returns following SEC events.

Our study contributes to the literature in several unique dimensions. Firstly, previous studies primarily rely on indirect measures like news coverage from specific agencies or Google Trends as proxies for regulatory action. Our paper advances previous research by constructing a precise dataset detailing each unexpected SEC enforcement action and each unexpected public announcement where crypto assets were classified as securities, leveraging the SEC's EDGAR database. Secondly, in contrast to previous studies, our dataset offers unprecedented detail by tracking each crypto asset explicitly classified as a security in each SEC announcement, enabling precise market impact analysis.[4] We include this data to facilitate future researchers. Thirdly, our empirical investigation employing a market model event study demonstrates a shift in momentum as CARs reverse from positive or relatively lower pre-announcement levels to negative following SEC classifications, intensifying in magnitude from -5.2% [0 to 3 days] to -17.2% [0 to 30 days]. In tandem, we reveal trading volumes sharply contract post-announcement, signaling a broader market retreat and potential investor exodus in response to SEC classifications.

---

[3] Our research focuses on the market's response to SEC regulatory announcements rather than attempting to forecast future price movements. While predictive models have value in assessing market trends, as noted in studies such as Gerritsen et al. (2022) and Meyer et al. (2024), our objective is to analyze the immediate and sustained effects of regulatory interventions.

[4] We identify 117 instances where the SEC classified blockchain-based tokens as securities, yet many still trade on U.S.-regulated exchanges like Coinbase and Kraken. This raises concerns about whether these actions align

Fourthly, our study identifies practical considerations for stakeholders. We investigate how market characteristics of assets named by the SEC affect their responses to being classified as securities. Larger market capitalization and older, more mature assets—traditional asset quality indicators—are not insulated from adverse effects of SEC classifications, challenging conventional market precepts. However, more illiquid assets face sharp post-announcement declines in returns, highlighting challenges in attracting buyers. Meanwhile, more volatile crypto assets experience a sharp decline, intensifying over a month, reflecting prolonged regulatory uncertainty. Positive sentiment during SEC announcements also plays a nuanced role, moderating, but not reversing, adverse effects. Lastly, we identify pre-announcement effects hinting at informed trading.

## 2 Data Background

The underlying dataset, detailed in Table 1, comprises event dates derived from the U.S. Securities and Exchange Commission (SEC) website (SEC, 2023a), documenting the classification of assets as securities. The classifications arise from enforcement actions or unexpected public announcements via the General Form for Registration of Securities (Form 10). We validated each event through systematic searches of press releases, news articles, litigation releases, and public statements targeting specific crypto assets.

**[Insert Table 1]**

Motivated by recent cryptocurrency research, we investigate the role of: '$\beta_1$(Size)', measured as log-transformed market capitalization, following research that crypto assets with larger market capitalizations are more resilient to shocks attributable to a broader investor base (Moratis, 2021); '$\beta_2$(Age), defined as the number of days each crypto asset has been listed as tradeable by CoinGecko relative to each event date[5], guided by findings that older, more mature assets engender investor trust and signal adoption (Cong et al., 2023); market-specific metrics, including '$\beta_3$(Volatility)', calculated over each estimation window, and '$\beta_4$(Illiquidity)' expressed as per (Amihud, 2002), inspired by

---

with the SEC's claims of investor protection and market integrity. Additionally, it is puzzling why the SEC permitted Coinbase's NASDAQ listing in April 2021, given that Coinbase facilitated trading in what the SEC considers unregulated securities. This suggests that the SEC's actions may not always adhere to its stated goals.

[5] Assets may have been tradeable over the counter (OTC) before listing on CoinGecko. However, utilizing CoinGecko data ensures equal treatment of all crypto assets.

research that assets with higher illiquidity require greater expected returns to compensate for risk (Leirvik, 2022), respond slower to news due to more expansive availability on decentralized exchanges (Hansen et al., 2024), and exhibit greater short-term effects (Brauneis et al., 2021); and '$\beta_5$(Sentiment)' gauged by the Alternative cryptocurrency sentiment index[6], aligning with literature demonstrating the influence of prevailing cryptocurrency sentiment on the market reaction intensity (Saggu, 2022; Wang et al., 2023).[7]

## 3 Empirical model and results

### 3.1 Baseline event-study model

We initiate the empirical investigation by applying an event study to evaluate market reactions to SEC classifications of specific crypto assets as securities. The economic significance of events is quantified by comparing log returns and log-transformed trading volumes of crypto assets in defined windows against expected returns derived from historical data (Boehmer et al., 1991). The analysis employs a market model (MacKinlay, 1997), with Bitcoin log returns and log-transformed trading volumes as the benchmark, to account for the potential influence of overarching market trends—consistent with related studies (e.g., Meyer and Ante, 2020).[8]

The market model can be expressed as $R_i = \alpha_i + \beta_i R_m + \epsilon_i$, where $R_i$ is the return (volume) of the crypto assets, $R_m$ is the market return (volume), $\alpha_i$ is the asset-specific intercept, $\beta_i$ measures the asset's sensitivity to market returns, and $\epsilon_i$ is the error term. We calculate expected returns (volumes) using a time series regression over an estimation period of -150 to -10 days before each event. The selection of a 140-day window first ensures potential pre-market reactions (i.e., anticipation effects and informed trading) can be analyzed, second prevents events from overlapping, overcoming complications relating to cross-sectional correlations (MacKinlay, 1997), and third ensures the length is adequate to overcome estimation sensitivity (Armitage, 1995).

Tables 2 and 3 present abnormal returns and abnormal volumes, respectively, using the market model for the (a) full sample period (Event IDs 1–48) and three sub-sample periods.[9] These include (b) the

---

[6] The Alternative.me Crypto Fear & Greed Index quantifies cryptocurrency sentiment on a scale from 0 ("Extreme Fear") to 100 ("Extreme Greed"). It integrates volatility, momentum/volume, social media, surveys, dominance, and Google Trends data. The index, updated daily, indicates buying opportunities when fear prevails and potential corrections during greed phases (Alternative.me, 2023).

[7] $\beta_2$(Age) is scaled by $10^2$ and $\beta_4$(Illiquidity) is scaled by $10^6$ for readability.

[8] In line with related studies, Bitcoin is used as our market proxy because it dominates over half of the total cryptocurrency market capitalization and serves as a barometer for the sector. Bitcoin's high trading volume and liquidity make it a reliable indicator for overarching market trends. By benchmarking each cryptocurrency against Bitcoin, we effectively isolate abnormal returns directly attributable to each event while controlling for broader market trends.

[9] The results are presented as graphs in Figures A.1. and A.2.

Binance (IDs 32–41, June 5, 2023) and Coinbase (IDs 42–47, June 6, 2023) events, isolating the impact of sustained regulatory scrutiny over two consecutive days; (c) the Coinbase insider trading event (IDs 5–13, July 21, 2022), and (d) the Bittrex enforcement action (IDs 26–31, April 17, 2023), confining reactions to simultaneous classifications of multiple assets on the exchange.

The immediate market reaction to SEC announcements across all events (Panel a) is significantly negative, with CARs deepening in magnitude from -5.2% [0,2 days], -12.2% [0,6], -13.5% [0,13] to a peak of -17.2% [0,30]. Our findings are consistent with the efficient market hypothesis, as markets rapidly integrate new regulatory news into the prices of assets identified by the SEC as security, promptly following announcements. The prolonged negative price adjustment process reflects a dynamic and evolving understanding of the complex and uncertain implications of regulatory changes as markets reevaluate affected assets, gradually incorporating the lack of subsequent SEC clarification into valuations. In stark contrast, the statistically insignificant and relatively subdued pre-announcement CARs of -2.4% indicate that the market had not anticipated the announcements, supporting efficient market incorporation of new public information.

In Panels (b) and (d), similar trends of deepening adverse reactions are evident post-announcement. Specifically, CARs in Panel (b) swiftly decline -6.5% [0,2], demonstrating the market's capacity to efficiently assimilate and respond to new information– to a peak of -22.6% [0,13], recovering slightly to -18.0% [0,30]. The positive and highly significant pre-announcement CARs of 2.5% [-7,-1] mark momentum reversal following SEC announcements, reinforcing the unexpected nature of the news and rapid reversal as markets efficiently assimilate new information. In contrast, Panel (c) shows no significant CARs post-announcement but a significant pre-announcement AR decline of -3.9% [-1 day], hinting at potential leaks or insider activity.

[Insert Table 2]

The impact on CAVs in Table 3 indicates a significant reduction in trading volume across panels (a) to (c), ranging from -4.95 to -10.85 units decline in log-transformed trading volume [0,30 days] post-announcement, underscoring a fall in demand for crypto assets in tandem with declining prices, or investors ceasing trading in respective assets. Interestingly, while CAVs steadily decline in panel (c) from -1.11 [0,2 days] to -10.85 [0,30], they increase in panel (d) from 1.47 [0,2] to a peak of 3.47 [0,13]. CAV estimates for shorter windows in panels (a) and (b) are less conclusive. While we broadly observe declines in CARs, with mixed results for CAVs, further analysis is required to determine the underlying factors determining the direction and magnitude of responses. In the context of efficient markets, investors may have adjusted positions based on SEC announcements or opted to refrain from trading, recognizing the rapid reflection of news in asset prices and thus perceiving fewer profitable opportunities.

**[Insert Table 3]**

## 3.2 Determinants of direction and magnitude

Building on our findings that CARs generally decline post-SEC announcements, while CAVs show mixed outcomes, we investigate the determinants influencing the magnitude of market reactions. The estimates in Table 4 employ robust MM estimator models to examine how the factors defined in Section 2 influence the CARs/AR and CAV/AV of cryptocurrencies classified as securities over different event windows and the event itself.[10] [11] The initial estimate (1), covering the one-week [-7,-1] window preceding each SEC announcement, reveals no significant explanatory power of the five factors for CARs/ARs, consistent with the efficient markets hypothesis (i.e., the absence of public information). This initial finding does not exclude the possibility of informed trading but indicates the factors do not account for pre-announcement variations in CARs/AR. In contrast, the second estimate (2) on the event day [0,0] reveals that SEC announcements significantly and negatively impact higher-volatility crypto assets ($\beta_3$=-0.765) with sharp announcement day declines. The negative effect intensifies over time, deepening ($\beta_3$=-5.151) in the one-month [0,30] window (estimate 6). Our findings are rooted in theories of risk aversion and information asymmetry. As SEC interventions heighten perceived regulatory risks, investors promptly sell assets to minimize losses driven by risk aversion amid uncertain environments. The effect is more pronounced in higher-volatility assets, where uncertainty further intensifies risk perceptions. Information asymmetry compounds the effect as lack of regulatory clarity makes it difficult for investors to assess the impact, driving overreactions and deeper sell-offs. The sustained reaction reflects ongoing market adjustments as investors process the regulatory uncertainty introduced by the SEC, consistent with studies that highlight the longer-term influence of news on cryptocurrency markets (Yue et al., 2021).

**[Insert Table 4]**

Three days [0-2] from each announcement (estimate 3), assets named by the SEC with higher illiquidity scores ($\beta_4$=-1.622) and larger market capitalization ($\beta_1$=-0.007) experience a fall in abnormal returns, indicating their vulnerability to regulatory news. In the context of liquidity premium theory, illiquid assets require higher returns to compensate for trading difficulty. As regulatory uncertainty from SEC announcements increases perceived risks, the premium widens, making it even harder for investors to find buyers without significantly impacting prices. As a result, illiquid assets experience sharper price declines. Signaling theory explains the smaller decline in larger market capitalization assets due to their role as industry bellwethers. Given their size and visibility, investors see these assets as regulatory targets. However, their broader investor base spreads the impact, muting the overall decline. The three-day market response lags are consistent with the time market participants need to respond to complex regulatory announcements, consistent with research identifying sluggish price discovery following news announcements in cryptocurrency markets (Hashemi Joo et al., 2020).

---

[10] MM estimates are more robust in the presence of significant and cluster outliers. We report the adjusted rw-squared statistic of Renaud and Victoria-Feser (2010) for these models.

[11] Descriptive statistics and correlations can be assessed in Tables A.1 and A.2.

Estimate (7) identifies statistically significant factors explaining variations in trading volumes during the one-week [-7,-1] window preceding each SEC announcement, hinting at informed trading, as investors potentially act on confidential information or rumors in anticipation of public disclosure.[12] Before each SEC announcement, assets with higher volatility ($\beta_3$=-8.534) and illiquidity ($\beta_4$=-0.410) due to be named in upcoming announcements experience a sharp fall in trading volumes. The results suggest that the perceived risk from upcoming regulatory announcements deter informed investors from trading in these assets, prompting strategic waiting until regulatory clarification. Furthermore, the lack of significant corresponding price movements in estimate (1) indicates that market depth and capacity to absorb new trades may not significantly impact prices. Despite potential preemptive selling by informed traders, price discovery in illiquid markets remains stable until public information is released and all market participants can respond.

Trading volumes for less liquid assets ($\beta_4$) substantially declined in the three days [0,2] following SEC announcements ($\beta_4$=-15.782), indicative of deferred reactions as markets assimilated complex regulatory changes. After two weeks [0,13], they intensified ($\beta_4$=-30.332), indicating deepening caution against less liquid assets amid the evaporation of liquidity in markets for already illiquid assets. The progressive drying up of liquidity may reflect withdrawal by buyers and sellers, traders holding back from transactions, and waiting for more definitive regulatory guidelines or market stabilization, compounded by the absence of clarification or further guidance from the SEC, intensifying the effect and discouraging re-engagement with the assets.

Lastly, estimates (7) to (12) highlight the crucial role of crypto sentiment ($\beta_5$) in significantly influencing trading volumes following events. Higher sentiment at the time of SEC announcements increases trading volumes as investors engage in speculative trading or strategic adjustments to capitalize on market movements. Conversely, lower sentiment suppresses trading activity as investors minimize losses or hold on to assets amidst adverse regulatory news and pessimistic sentiment. Coefficient magnitudes indicate that sentiment tempers rather than negates the adverse effects of SEC announcements.

---

[12] Although our study observes market patterns indicative of pre-announcement activity, we cannot definitively attribute this behavior to informed trading. Subsequent studies may more rigorously assess the presence of informed trading in response to SEC regulatory interventions by employing trade-level analysis and order size imbalances, as demonstrated in Feng et al. (2018).

# 4  Conclusion

Our study brings to light the significant uncertainties and financial risks that cryptocurrency investors face in the aftermath of the SEC's unexpected regulatory interventions. Furthermore, the crypto industry is grappling with ambiguous regulatory frameworks that impede the establishment of fair, orderly, and efficient markets. In this context, the SEC's potential failure to uphold its core mandate of protecting investors and ensuring efficient markets (SEC, 2023b) is evident. The lack of clear guidelines for the cryptocurrency industry and the indiscriminate targeting of the sector aligns with Stigler's (1974) theory of regulatory capture, as the SEC—potentially influenced by incumbent financial entities—may favor established institutions over emerging technologies, thereby stifling innovation in the evolving crypto market. The public interest theory lens (Posner, 1974) reveals that severe adverse market reactions to SEC announcements highlight a counterproductive regulatory strategy that increases systemic risk, particularly for illiquid and volatile assets.

To address these challenges, the SEC could consider implementing a safe harbor provision, as proposed by the SEC's Hester Pierce (Melinek, 2024), which would allow crypto projects time to develop without being prematurely classified as securities. Additionally, expanding a regulatory sandbox could foster innovation while maintaining investor protection. A review of the decades-old Howey Test, used to determine whether an asset qualifies as a security, would also be beneficial. Providing clear guidelines on what constitutes a security would give the industry much-needed clarity and help prevent undue targeting. Without transparent criteria, the SEC risks falling short of its mandate to protect investors and maintain orderly, efficient markets. Future research should explore potential informed trading or information leaks from the SEC and analyze the long-term effects of regulatory (non-)interventions to assess broader market stability and provide deeper insights into the optimization of regulatory frameworks to support innovation while safeguarding investor interests and maintaining market integrity. Expanding the analysis to cover a more diverse range of crypto assets would further strengthen the generalizability of the results

**Table 1. Overview of SEC classifications of individual crypto assets as securities**

Note: Table 1 details events where the United States Securities and Exchange Commission (SEC) classified individual crypto assets as securities, communicated through enforcement actions or unexpected public announcements via General Form for Registration of Securities (Form 10). Each row presents a serial number for the sequential asset-level event (#), the asset ticker code, the full asset name, and a reference to the SEC documentation substantiating the classification event. The event identification (Event ID) column lists events meeting the study's inclusion criteria: (i) available daily trading data from CoinGecko and (ii) consideration of overlaps pertinent to classifications.

| # | Event ID | Date (dd/mm/yyyy) | Asset Ticker | Asset Name | Reference Documentation |
|---|---|---|---|---|---|
| 1 | | 25/07/2017 | DAO | The DAO | SEC regulatory action |
| 2 | | 29/09/2017 | REC | REcoin | SEC regulatory action |
| 3 | | 29/09/2017 | DRC | DRC World | SEC regulatory action |
| 4 | | 01/12/2017 | PLEX | PlexCoin | SEC regulatory action |
| 5 | | 11/12/2017 | MUN | Munchee | SEC regulatory action |
| 6 | | 25/01/2018 | ACO | AriseCoin | SEC regulatory action |
| 7 | | 20/03/2018 | BAR | Titanium | SEC regulatory action |
| 8 | | 14/08/2018 | TOM | Tomahawkcoins | SEC regulatory action |
| 9 | | 16/11/2018 | AIR | Airtoken | SEC regulatory action |
| 10 | | 16/11/2018 | PRG | Paragon | SEC regulatory action |
| 11 | | 29/11/2018 | CTR | Centra | SEC regulatory action |
| 12 | | 20/02/2019 | GLA | Gladius | SEC regulatory action |
| 13 | 1 | 04/06/2019 | KIN | KIN | SEC regulatory action |
| 14 | | 12/08/2019 | VERI | Veritaseum | SEC regulatory action |
| 15 | | 12/08/2019 | HLTH | Health Cash | SEC regulatory action |
| 16 | | 29/08/2019 | Bitqy | Bitqyck | SEC regulatory action |
| 17 | | 29/08/2019 | BitqyM | Bitqyck Mining | SEC regulatory action |
| 18 | | 18/09/2019 | ICOS | ICO Box | SEC regulatory action |
| 19 | | 20/09/2019 | FM | Fantasy Market | SEC regulatory action |
| 20 | | 30/09/2019 | EOS | EOS | SEC regulatory action |
| 21 | | 11/10/2019 | TON | Toncoin | SEC regulatory action |
| 22 | | 11/12/2019 | SHOP | Shopin | SEC regulatory action |
| 23 | | 18/12/2019 | BCOT | Blockchain of Things | SEC regulatory action |
| 24 | | 21/01/2020 | OPP | OPP Tokens | SEC regulatory action |
| 25 | | 19/02/2020 | ENG | Enigma | SEC regulatory action |
| 26 | | 27/02/2020 | B2G | Bitcoiin2Gen | SEC regulatory action |
| 27 | | 20/03/2020 | META1 | Meta 1 Coin | SEC regulatory action |
| 28 | | 24/04/2020 | DROP | Dropil | SEC regulatory action |
| 29 | | 28/05/2020 | CAT | Consumer Activity Token | SEC regulatory action |
| 30 | | 13/08/2020 | BOON | Boon Coins | SEC regulatory action |
| 31 | | 11/09/2020 | SPARK | CoinSpark | SEC regulatory action |
| 32 | | 11/09/2020 | FLiK | FliK | SEC regulatory action |
| 33 | | 15/09/2020 | UKG | UniKoinGold | SEC regulatory action |
| 34 | | 25/09/2020 | XD | Scroll Network | SEC regulatory action |
| 35 | 2 | 30/09/2020 | SALT | Salt | SEC regulatory action |
| 36 | | 09/12/2020 | PEARL | Oyster Protocol | SEC regulatory action |
| 37 | | 21/12/2020 | SHIP | ShipChain | SEC regulatory action |
| 38 | 3 | 22/12/2020 | XRP | Ripple / XRP | SEC regulatory action |
| 39 | | 23/12/2020 | TNT | Tierion | SEC regulatory action |
| 40 | | 09/01/2021 | BCC | BitConnect | SEC regulatory action |
| 41 | | 15/01/2021 | WRL | Wireline | SEC regulatory action |
| 42 | | 01/02/2021 | B2G | Bitcoiin2Gen | SEC regulatory action |
| 43 | 4 | 29/03/2021 | LBC | LBRY Credits | SEC regulatory action |
| 44 | | 15/06/2021 | BCT | BCT Tokens | SEC regulatory action |
| 45 | | 22/06/2021 | LOCI | LOCIcoin | SEC regulatory action |
| 46 | | 04/08/2021 | UULA | Uulala | SEC regulatory action |
| 47 | | 06/08/2021 | DMG | DeFi Money Market | SEC regulatory action |
| 48 | | 06/08/2021 | mToken | DeFi Money Market | SEC regulatory action |
| 49 | | 08/09/2021 | RVT | Rivetz | SEC regulatory action |
| 50 | | 02/12/2021 | DNO | Denaro | SEC regulatory action |
| 51 | | 06/01/2022 | CMCT | Crowd Machine C. T. | SEC regulatory action |
| 52 | | 08/03/2022 | ORV | Ormeus Coin | SEC regulatory action |
| 53 | | 28/04/2022 | NSG | NASGO | SEC regulatory action |
| 54 | | 28/04/2022 | SNP | Sharenode | SEC regulatory action |
| 55 | 5 | 21/07/2022 | AMP | Amp | SEC regulatory action |
| 56 | 6 | 21/07/2022 | POWR | Power Ledger | SEC regulatory action |
| 57 | 7 | 21/07/2022 | RLY | Rally | SEC regulatory action |

| # | # | Date | Symbol | Name | Link |
|---|---|------|--------|------|------|
| 58 | 8 | 21/07/2022 | DDX | DerivaDAO | SEC regulatory action |
| 59 | 9 | 21/07/2022 | XYO | XYO Network | SEC regulatory action |
| 60 | 10 | 21/07/2022 | RGT | Rari Governance Token | SEC regulatory action |
| 61 | 11 | 21/07/2022 | LCX | L. Cryptoasset Exchange | SEC regulatory action |
| 62 | 12 | 21/07/2022 | DFX | DFX Finance | SEC regulatory action |
| 63 | 13 | 21/07/2022 | KROM | Kromatica | SEC regulatory action |
| 64 | 14 | 16/08/2022 | DRGN | DragonChain | SEC regulatory action |
| 65 | | 08/09/2022 | BLT | Bloom | SEC regulatory action |
| 66 | | 14/09/2022 | BXY | Beaxy | SEC regulatory action |
| 67 | | 19/09/2022 | SPRK | Sparkster | SEC regulatory action |
| 68 | 15 | 29/09/2022 | HYDRO | Hydro | SEC regulatory action |
| 69 | | 29/09/2022 | DIG | Dignity | SEC regulatory action |
| 70 | 16 | 03/10/2022 | EMAX | Ethereum Max | SEC regulatory action |
| 71 | | 18/11/2022 | DUCAT | Ducat | SEC regulatory action |
| 72 | | 18/11/2022 | LOCKE | Locke | SEC regulatory action |
| 73 | 17 | 21/12/2022 | FTT | FTX Token | SEC regulatory action |
| 74 | 18 | 19/01/2023 | NEXO | Nexo | SEC regulatory action |
| 75 | 19 | 20/01/2023 | MNGO | Mango | SEC regulatory action |
| 76 | 20 | 16/02/2023 | LUNA | Luna | SEC regulatory action |
| 77 | 21 | 16/02/2023 | MIR | Mirror Protocol | SEC regulatory action |
| 78 | 22 | 16/02/2023 | UST | Terra USD | SEC regulatory action |
| 79 | | 17/02/2023 | EMAX | Ethereum Max | SEC regulatory action |
| 80 | | 08/03/2023 | GREEN | Green | SEC regulatory action |
| 81 | 23 | 17/03/2023 | FIL | Filecoin | SEC regulatory action |
| 82 | 24 | 22/03/2023 | BTT | Bittorrent | SEC regulatory action |
| 83 | 25 | 22/03/2023 | TRX | TRON | SEC regulatory action |
| 84 | | 29/03/2023 | BXY | Beaxy Token | SEC regulatory action |
| 85 | 26 | 17/04/2023 | ALGO | Algorand | SEC regulatory action |
| 86 | 27 | 17/04/2023 | DASH | Dash | SEC regulatory action |
| 87 | 28 | 17/04/2023 | IHT | I-House Token | SEC regulatory action |
| 88 | 29 | 17/04/2023 | NGC | NAGA | SEC regulatory action |
| 89 | 30 | 17/04/2023 | OMG | OmiseGo | SEC regulatory action |
| 90 | 31 | 17/04/2023 | TKN | Monolith | SEC regulatory action |
| 91 | | 18/04/2023 | UP | UpToken | SEC regulatory action |
| 92 | 32 | 05/06/2023 | ADA | Cardano | SEC regulatory action |
| 93 | | 05/06/2023 | ALGO | Algorand | SEC regulatory action |
| 94 | 33 | 05/06/2023 | ATOM | Cosmos Hub | SEC regulatory action |
| 95 | 34 | 05/06/2023 | AXS | Axie Infinity | SEC regulatory action |
| 96 | 35 | 05/06/2023 | BNB | Binance Coin | SEC regulatory action |
| 97 | 36 | 05/06/2023 | BUSD | Binance USD | SEC regulatory action |
| 98 | 37 | 05/06/2023 | COTI | Coti | SEC regulatory action |
| 99 | | 05/06/2023 | FIL | Filecoin | SEC regulatory action |
| 100 | 38 | 05/06/2023 | MANA | Decentraland | SEC regulatory action |
| 101 | 39 | 05/06/2023 | MATIC | Polygon | SEC regulatory action |
| 102 | 40 | 05/06/2023 | SAND | The Sandbox | SEC regulatory action |
| 103 | 41 | 05/06/2023 | SOL | Solana | SEC regulatory action |
| 104 | 42 | 06/06/2023 | VGX | Voyager Token | SEC regulatory action |
| 105 | | 06/06/2023 | SOL | Solana | SEC regulatory action |
| 106 | 43 | 06/06/2023 | NEXO | Nexo | SEC regulatory action |
| 107 | 44 | 06/06/2023 | NEAR | Near | SEC regulatory action |
| 108 | 45 | 06/06/2023 | ICP | Internet Computer | SEC regulatory action |
| 109 | 46 | 06/06/2023 | FLOW | Flow | SEC regulatory action |
| 110 | 47 | 06/06/2023 | CHZ | Chiliz | SEC regulatory action |
| 111 | | 06/06/2023 | ADA | Cardano | SEC regulatory action |
| 112 | | 06/06/2023 | FIL | Filecoin | SEC regulatory action |
| 113 | | 06/06/2023 | MATIC | Polygon | SEC regulatory action |
| 114 | | 06/06/2023 | SAND | The Sandbox | SEC regulatory action |
| 115 | | 06/06/2023 | AXS | Axie Infinity | SEC regulatory action |
| 116 | | 06/06/2023 | DASH | Dash | SEC regulatory action |
| 117 | 48 | 13/07/2023 | CEL | Celsius Network | SEC regulatory action |

**Table 2. Event study results for crypto asset log returns**

Note: Table 2 presents event study results for crypto asset log returns, using a market model with Bitcoin as a benchmark market and a 140-day estimation period. The analysis covers 48 U.S. Securities and Exchange Commission (SEC) events classifying cryptocurrencies as securities in Panel (a), and sub-samples in Panels (b) for Binance and Coinbase events, (c) for Coinbase insider trading, and (d) for Bittrex enforcement actions. Estimates include cumulative abnormal returns (CARs) and individual abnormal returns (ARs) assessed via t-tests and the non-parametric Wilcoxon (1945) sign rank test ('z-test'), with significance levels indicated by *, **, and *** for 10%, 5%, and 1% levels, respectively.

|  | **Panel (a)** **All events** | | | **Panel (b)** **Binance and Coinbase** | | | **Panel (c)** **Coinbase Insider Trading** | | | **Panel (d)** **Bittrex Enforcement** | | |
|---|---|---|---|---|---|---|---|---|---|---|---|---|
| Window | CAR | t-test | z-test | CAR | t-test | z-test | CAR | t-test | z-test | CAR | t-test | z-test |
| [-7, -1] | -0.024 | -1.95** | -1.34 | 0.025 | 3.78*** | 3.05*** | -0.035 | -1.00 | -0.89 | -0.036 | -1.86* | -1.57 |
| [0, 2] | -0.052 | -1.80* | -3.73*** | -0.065 | -7.72*** | -3.41*** | -0.052 | -1.09 | -1.24 | -0.044 | -1.47 | -0.94 |
| [0, 6] | -0.122 | -4.12*** | -4.40*** | -0.227 | -9.21*** | -3.52*** | -0.025 | -0.59 | -0.41 | -0.093 | -3.33*** | -2.20** |
| [0, 13] | -0.135 | -2.88*** | -3.67*** | -0.226 | -7.91*** | -3.52*** | -0.004 | -0.04 | 0.06 | -0.231 | -1.84* | -1.99** |
| [0, 30] | -0.172 | -3.61** | -3.39*** | -0.180 | -4.59*** | -2.95*** | 0.008 | 0.12 | 0.30 | -0.252 | -2.41** | -1.99** |
| Day | AR | t-test | z-test | AR | t-test | z-test | AR | t-test | z-test | AR | t-test | z-test |
| [-7] | -0.009 | -1.01 | -2.79*** | -0.008 | -1.96** | -1.96** | 0.015 | 0.40 | -1.01 | 0.000 | 0.02 | -0.94 |
| [-6] | -0.016 | -3.00*** | -3.23*** | 0.000 | 0.09 | -0.05 | -0.002 | -0.14 | -1.01 | -0.020 | -1.54 | -1.36 |
| [-5] | -0.000 | -0.04 | -0.51 | 0.006 | 1.43 | 1.03 | -0.008 | -1.10 | -0.65 | -0.041 | -3.66*** | -2.20** |
| [-4] | 0.001 | 0.14 | -0.25 | 0.006 | 1.67* | 1.45 | 0.024 | 1.01 | 0.77 | -0.041 | -2.45** | -1.99** |
| [-3] | 0.005 | 0.64 | 0.09 | 0.012 | 2.38** | 2.12** | -0.023 | -2.87*** | -2.07** | 0.040 | 1.32 | 1.99** |
| [-2] | 0.006 | 1.05 | 1.34 | 0.008 | 3.05*** | 2.59*** | -0.002 | -0.16 | -0.65 | 0.026 | 3.13*** | 2.20** |
| [-1] | -0.012 | 1.61 | -2.07** | 0.001 | 0.49 | 0.98 | -0.039 | -2.17** | -2.67*** | -0.000 | -0.00 | 0.31 |
| [0] | -0.011 | -1.92* | -1.63 | -0.003 | -0.48 | -0.21 | -0.023 | -1.85* | -1.60 | 0.016 | 0.96 | 1.15 |
| [1] | -0.016 | -0.67 | -3.21*** | -0.025 | -4.56 | -3.31 | -0.027 | -1.67* | -1.13 | -0.047 | -1.20 | -1.36 |
| [2] | -0.024 | -1.68* | -2.22** | -0.038 | -7.01 | -3.46 | -0.002 | -0.07 | -0.06 | -0.013 | -1.76* | -1.57 |
| [3] | -0.010 | -0.99 | -1.89* | -0.028 | -3.65 | -2.90 | 0.021 | 1.13 | 1.01 | -0.028 | -2.09** | -1.78* |
| [4] | -0.014 | -2.18** | -2.04** | -0.004 | -1.11 | -1.03 | 0.005 | 0.57 | 0.65 | -0.034 | -1.73* | -2.20** |
| [5] | -0.031 | -3.55*** | -3.37*** | -0.071 | -4.34 | -3.36 | -0.010 | -0.53 | -0.41 | -0.015 | -2.09** | -1.57 |
| [6] | -0.015 | -1.45 | -1.88* | -0.059 | -3.43 | -2.79 | 0.011 | 0.69 | 0.89 | 0.028 | 1.07 | 1.15 |

## Table 3. Event study results for crypto asset log volumes

Table 3 presents event study results for crypto asset log volumes, using a market model with Bitcoin as a benchmark market and a 140-day estimation period. The analysis covers 48 U.S. Securities and Exchange Commission (SEC) events classifying cryptocurrencies as securities in Panel (a), and sub-samples in Panels (b) for Binance and Coinbase events, (c) for Coinbase insider trading, and (d) for Bittrex enforcement actions. Estimates include cumulative abnormal trading volumes (CAVs) and mean abnormal trading volumes (AVs) assessed via t-tests and the non-parametric Wilcoxon (1945) sign rank test ('z-test'), with significance levels indicated by *, **, and *** for 10%, 5%, and 1% levels, respectively.

| | Panel (a) All events | | | Panel (b) Binance and Coinbase | | | Panel (c) Coinbase Insider Trading | | | Panel (d) Bittrex Enforcement | | |
|---|---|---|---|---|---|---|---|---|---|---|---|---|
| Window | CAR | $t$-test | $z$-test | CAR | $t$-test | $z$-test | CAR | $t$-test | $z$-test | CAR | $t$-test | $z$-test |
| [-7, -1] | -1.230 | -1.84* | -1.84* | -1.855 | -3.25*** | -3.00*** | -3.240 | -1.95* | -1.72* | 3.015 | 2.82*** | 1.99** |
| [0, 2] | 0.227 | 0.69 | 0.28 | -0.250 | -0.20 | -0.26 | -1.114 | -3.16*** | -2.19** | 1.467 | 4.25*** | 2.20** |
| [0, 6] | 0.334 | 0.52 | 0.81 | 0.396 | 0.57 | 0.98 | -2.741 | -3.01*** | -2.31** | 2.239 | 3.31*** | 2.20** |
| [0, 13] | -0.359 | -0.31 | 0.13 | -0.540 | -0.39 | -0.16 | -4.029 | -2.33** | -1.83* | 3.474 | 2.21** | 1.99** |
| [0, 30] | -4.952 | -1.92* | -2.03** | -5.971 | -2.14** | -2.17** | -10.848 | -2.74** | -2.07** | 1.377 | 0.23 | -0.31 |
| Day | AR | $t$-test | $z$-test | AR | $t$-test | $z$-test | AR | $t$-test | $z$-test | AR | $t$-test | $z$-test |
| [-7] | -0.190 | -1.55 | -1.91* | -0.319 | -2.96*** | -2.48** | -0.325 | -0.77 | -1.60 | 0.333 | 1.97** | 1.57 |
| [-6] | -0.270 | -2.05** | -2.59** | -0.390 | -3.61*** | -2.69*** | -0.334 | -0.80 | -1.60 | 0.494 | 2.09** | 1.57 |
| [-5] | -0.277 | -2.19** | -2.45** | -0.187 | -2.42** | -2.02** | -0.702 | -1.86* | -1.72* | 0.442 | 1.55 | 1.57 |
| [-4] | -0.094 | -0.77 | -1.06 | -0.113 | -1.23 | -1.19 | -0.28 | -0.89 | -1.13 | 0.581 | 2.05** | 1.99** |
| [-3] | -0.106 | -1.03 | -0.93 | -0.256 | -2.09** | -1.91** | -0.336 | -1.71* | -1.72* | 0.430 | 2.16** | 1.78** |
| [-2] | -0.099 | -0.91 | -1.28 | -0.344 | -3.26*** | -2.69*** | -0.663 | -3.70*** | -2.31** | 0.359 | 3.73*** | 2.20** |
| [-1] | -0.193 | -1.67* | -1.72* | -0.245 | -1.84* | -1.65* | -0.595 | -3.27*** | -2.31** | 0.375 | 3.39*** | 2.20** |
| [0] | -0.027 | -0.28 | 0.25 | 0.016 | 0.13 | -0.16 | -0.265 | -1.86* | -1.60 | 0.248 | 1.68* | 1.57 |
| [1] | 0.114 | 0.78 | 0.25 | -0.067 | -0.41 | -0.62 | -0.633 | -2.95** | -2.19** | 0.851 | 3.41*** | 2.20** |
| [2] | 0.139 | 0.94 | 0.28 | -0.199 | -1.86* | -1.76* | -0.217 | -1.01 | -1.01 | 0.369 | 3.14*** | 1.99** |
| [3] | 0.119 | 0.98 | 1.07 | -0.053 | -0.41 | 0.26 | -0.276 | -0.91 | -1.36 | 0.322 | 3.54*** | 2.20** |
| [4] | -0.095 | -1.05 | -0.62 | -0.188 | -1.58 | -1.60 | -0.489 | -2.22* | -1.60 | 0.300 | 2.06** | 1.57 |
| [5] | 0.026 | 0.23 | 0.40 | 0.278 | 1.82* | 1.65* | -0.173 | -0.50 | -1.36 | 0.060 | 0.44 | 0.52 |
| [6] | 0.058 | 0.46 | 0.54 | 0.608 | 3.07*** | 2.22** | -0.689 | -3.87*** | -2.43** | -0.004 | -0.02 | -0.10 |

**Table 4. Determinants of cumulative abnormal returns and trading volumes across different time windows.**

Note: Table 4 presents robust mm estimator regression models to analyze the determinants of cumulative abnormal returns (CARs) and cumulative abnormal trading volumes (CAVs) across different windows surrounding 48 U.S. Securities and Exchange Commission (SEC) events classifying cryptocurrencies as securities. $\beta_1$(Size) denotes each asset's log-transformed market capitalization. $\beta_2$ (Age) denotes the days CoinGecko has listed each crypto asset as tradeable, calculated relative to the date of each SEC event, and is scaled by $10^2$ for readability. $\beta_3$(Volatility) denotes the mean volatility of each asset over the estimation window. $\beta_4$(Illiquidity) denotes each asset's Amihud (2002) illiquidity metric and is scaled by $10^6$ for readability. $\beta_5$(Sentiment) denotes the Alternative.me cryptocurrency sentiment index on each event date. The table reports the adjusted rw-squared statistic of Renaud and Victoria-Feser (2010). Significance is indicated by *, **, and *** for 10%, 5%, and 1% levels, respectively.

|  | Returns | | | | | | Trading Volumes | | | | | |
|---|---|---|---|---|---|---|---|---|---|---|---|---|
|  | 1 Week Pre-event | 0 Day Event | 3 Day Post-event | 1 Week Post-event | 2 Week Post-event | 1 Month Post-event | 1 Week Pre-event | 0 Day Event | 3 Day Post-event | 1 Week Post-event | 2 Week Post-event | 1 Month Post-event |
| Estimation | (1) | (2) | (3) | (4) | (5) | (6) | (7) | (8) | (9) | (10) | (11) | (12) |
| $\beta_1$ (Size) | 0.002 (0.004) | -0.001 (0.003) | -0.007** (0.003) | -0.002 (0.011) | -0.011 (0.010) | -0.007 (0.022) | -0.243* (0.134) | 0.007 (0.019) | 0.077 (0.123) | 0.293** (0.121) | 0.326 (0.263) | 1.159* (0.645) |
| $\beta_2$ (Age) | 0.001 (0.002) | 0.001 (0.009) | 0.001 (0.001) | 0.002 (0.005) | 0.002 (0.005) | -0.004 (0.009) | -0.003 (0.009) | -0.002 (0.012) | -0.036 (0.064) | -0.001 (0.001) | -0.002 (0.001) | -0.001 (0.037) |
| $\beta_3$ (Volatility) | 0.255 (0.494) | -0.765*** (0.249) | -0.279 (0.313) | -0.248 (1.230) | -1.035 (0.122) | -5.151** (1.658) | -8.534 (18.843) | 1.949 (4.016) | 9.793 (10.487) | 5.599 (13.773) | -21.289 (22.502) | -63.314 (67.711) |
| $\beta_4$ (Illiquidity) | -0.001 (0.018) | 0.004 (0.004) | -1.622*** (0.039) | 0.010 (0.013) | 0.010 (0.008) | -0.044 (0.090) | -0.410** (0.175) | -0.009 (0.064) | -15.782** (6.400) | -18.995*** (3.612) | -30.332*** (9.889) | -424.158*** (137.880) |
| $\beta_5$ (Sentiment) | -0.001 (0.001) | 0.001 (0.001) | -0.002** (0.001) | -0.003 (0.002) | -0.003 (0.003) | -0.009** (0.004) | 0.178*** (0.044) | 0.012** (0.006) | 0.116*** (0.033) | 0.191*** (0.040) | 0.274*** (0.074) | 0.394** (0.178) |
| $\alpha_1$ (Constant) | -0.004 (0.114) | 0.031 (0.082) | 0.148* (0.083) | 0.061 (0.316) | 0.280 (0.271) | 0.713 (0.588) | -4.926 (3.564) | -0.796* (0.395) | -7.791** (3.341) | -14.180*** (3.254) | -17.733 (6.991) | -47.894** (19.370) |
| Adj. $Rw^2$ | 0.02 | 0.28 | 0.06 | 0.05 | 0.07 | 0.21 | 0.33 | 0.10 | 0.36 | 0.29 | 0.20 | 0.08 |
| Period | [-7,-1] | [0,0] | [0,2] | [0,6] | [0,13] | [0,30] | [-7,-1] | [0,0], | [0,2] | [0,6] | [0,13] | [0,30] |
| Dep. var. | CAR | AR | CAR | CAR | CAR | CAR | CAV | AV | CAV | CAV | CAV | CAV |

# Appendix

### Table A.1. Descriptive statistics

Note: Appendix Table A.1 present descriptive statistics for the variables used in the analysis of cryptocurrencies classified as securities by the U.S. Securities and Exchange Commission (SEC). 'Size' is measured as each asset's log-transformed market capitalization. 'Age' is measured as the days CoinGecko has listed each crypto asset as tradeable, calculated relative to the date of each SEC event, and is scaled by $10^2$ for readability. 'Volatility' is measured as the mean volatility of each asset over the estimation window. 'Illiquidity' is measured as each asset's Amihud (2002) illiquidity metric and is scaled by $10^6$ for readability. 'Sentiment' is measured as the Alternative.me cryptocurrency sentiment index on each event date. Test statistics include mean, standard deviation (SD), median, minimum (Min), and maximum (Max) values for each variable.

| Variables    | Mean   | SD     | Median | Min    | Max    |
|--------------|--------|--------|--------|--------|--------|
| Size         | 19.05  | 2.89   | 19.34  | 11.96  | 24.60  |
| Age          | 1,345  | 658    | 1,375  | 245    | 3,349  |
| Volatility   | 0.061  | 0.030  | 0.057  | 0.002  | 0.157  |
| Illiquidity  | -0.057 | 1.441  | 0.000  | -7.656 | 6.223  |
| Sentiment    | 51.44  | 14.77  | 54.00  | 22.00  | 88.00  |

### Table A.2. Correlations.

Note: Appendix Table A.2 presents the correlations among the variables and their impacts on cumulative abnormal returns (CARs) and cumulative abnormal trading volumes (CAVs) across various event windows surrounding U.S. Securities and Exchange Commission (SEC) announcements. The table delineates correlations within control variables and their respective influence on financial metrics across multiple timeframes, specified from one week pre-event to one-month post-event. Correlations and impact estimates are presented separately for control variables and their relationship with CARs and CAVs. 'Size' is measured as each asset's log-transformed market capitalization. 'Age' is measured as the days CoinGecko has listed each crypto asset as tradeable, calculated relative to the date of each SEC event, and is scaled by $10^2$ for readability. 'Volatility' is measured as the mean volatility of each asset over the estimation window. 'Illiquidity' is measured as each asset's Amihud (2002) illiquidity metric and is scaled by $10^6$ for readability. 'Sentiment' is measured as the Alternative.me cryptocurrency sentiment index on each event date. Significance is indicated by * for the 5% level.

| Variable     | Size    | Age     | Volatility | Amihud  | Sentiment |
|--------------|---------|---------|------------|---------|-----------|
| **(a) Controls** |     |         |            |         |           |
| Size         | 1.000   |         |            |         |           |
| Age          | 0.211   | 1.000   |            |         |           |
| Volatility   | -0.623* | -0.357* | 1.000      |         |           |
| Illiquidity  | 0.580*  | 0.210   | -0.374*    | 1.000   |           |
| Sentiment    | 0.345*  | 0.490*  | -0.327*    | 0.406*  | 1.000     |
| **(b) CARs** |         |         |            |         |           |
| [-7, -1]     | -0.013  | -0.042  | -0.012     | -0.133  | -0.081    |
| [0, 0]       | 0.223   | 0.278   | -0.340*    | 0.199   | 0.338*    |
| [0, 2]       | -0.015  | 0.025   | -0.211     | -0.056  | -0.205    |
| [0, 6]       | -0.269  | -0.085  | 0.114      | -0.192  | -0.225    |
| [0, 13]      | -0.220  | -0.096  | 0.013      | -0.251  | -0.303*   |
| [0, 30]      | -0.128  | -0.053  | -0.049     | -0.384* | -0.411*   |
| **(c) CAVs** |         |         |            |         |           |
| [-7, -1]     | 0.094   | 0.340*  | -0.183     | -0.067  | 0.497*    |
| [0, 0]       | 0.155   | 0.219   | -0.085     | 0.070   | 0.385*    |
| [0, 2]       | 0.034   | 0.268   | -0.099     | 0.129   | 0.451*    |
| [0, 6]       | 0.161   | 0.301*  | -0.180     | 0.175   | 0.441*    |
| [0, 13]      | 0.160   | 0.264   | -0.238     | 0.156   | 0.415*    |
| [0, 30]      | 0.139   | 0.127   | -0.233     | 0.028   | 0.225     |

**Figure A.1. Event study results for crypto asset log returns and log volumes [0, 30 days]**

Note: Figure A.1. presents event study results for crypto asset log returns, using a market model with Bitcoin as a benchmark market and a 140-day estimation period, over the first 30 days after each event. The analysis covers 48 U.S. Securities and Exchange Commission (SEC) events classifying cryptocurrencies as securities in Panel (a), and sub-samples in Panels (b) for Binance and Coinbase events, (c) for Coinbase insider trading, and (d) for Bittrex enforcement actions. The shaded gray areas denote 90% confidence intervals.

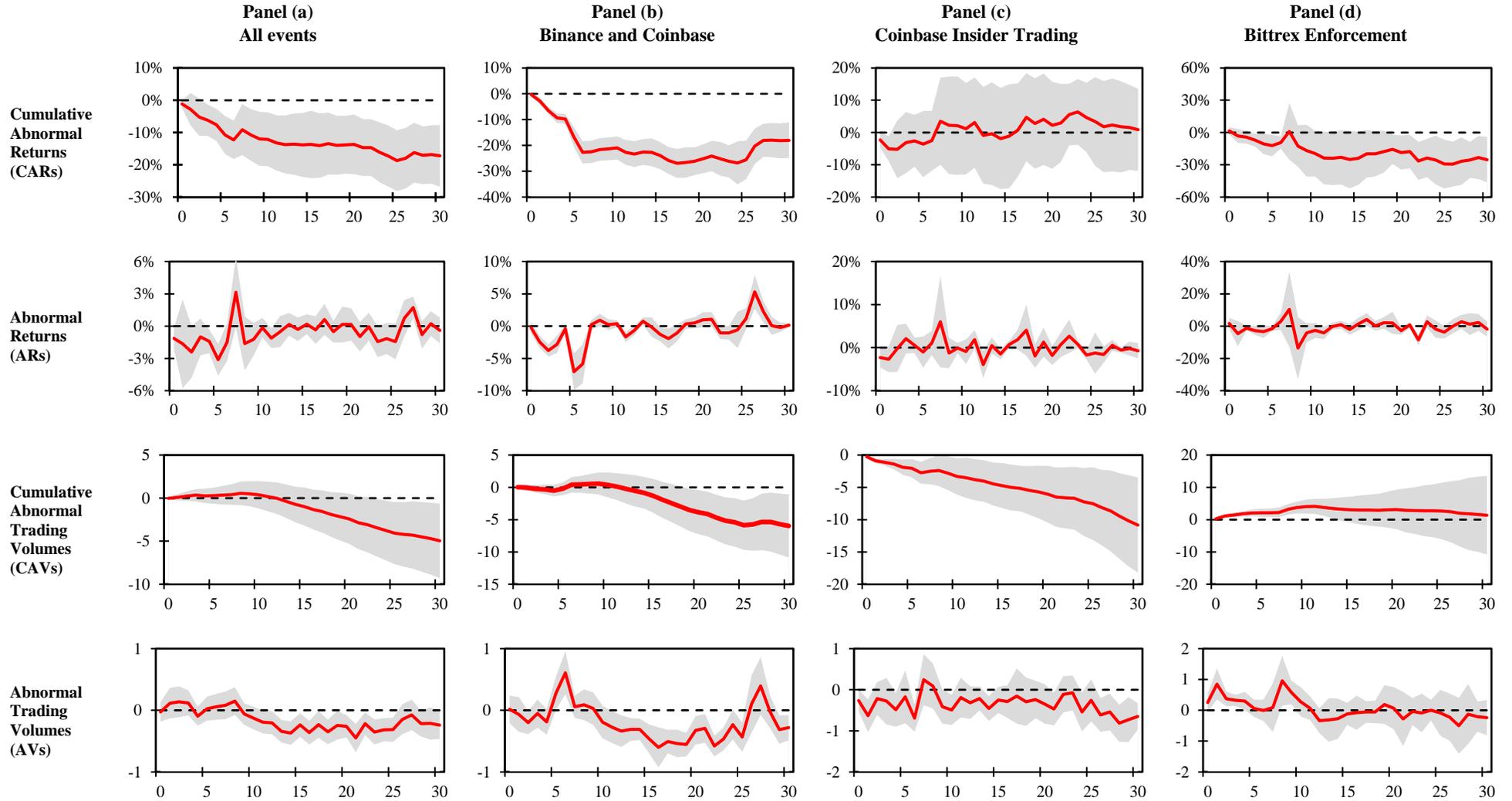

**Figure A.2. Event study results for crypto asset log returns and log volumes [-7, -1 days]**

Note: Figure A.2. presents event study results for crypto asset log returns, using a market model with Bitcoin as a benchmark market and a 140-day estimation period, over the 7 days preceding each event. The analysis covers 48 U.S. Securities and Exchange Commission (SEC) events classifying cryptocurrencies as securities in Panel (a), and sub-samples in Panels (b) for Binance and Coinbase events, (c) for Coinbase insider trading, and (d) for Bittrex enforcement actions. The shaded gray areas denote 90% confidence intervals.

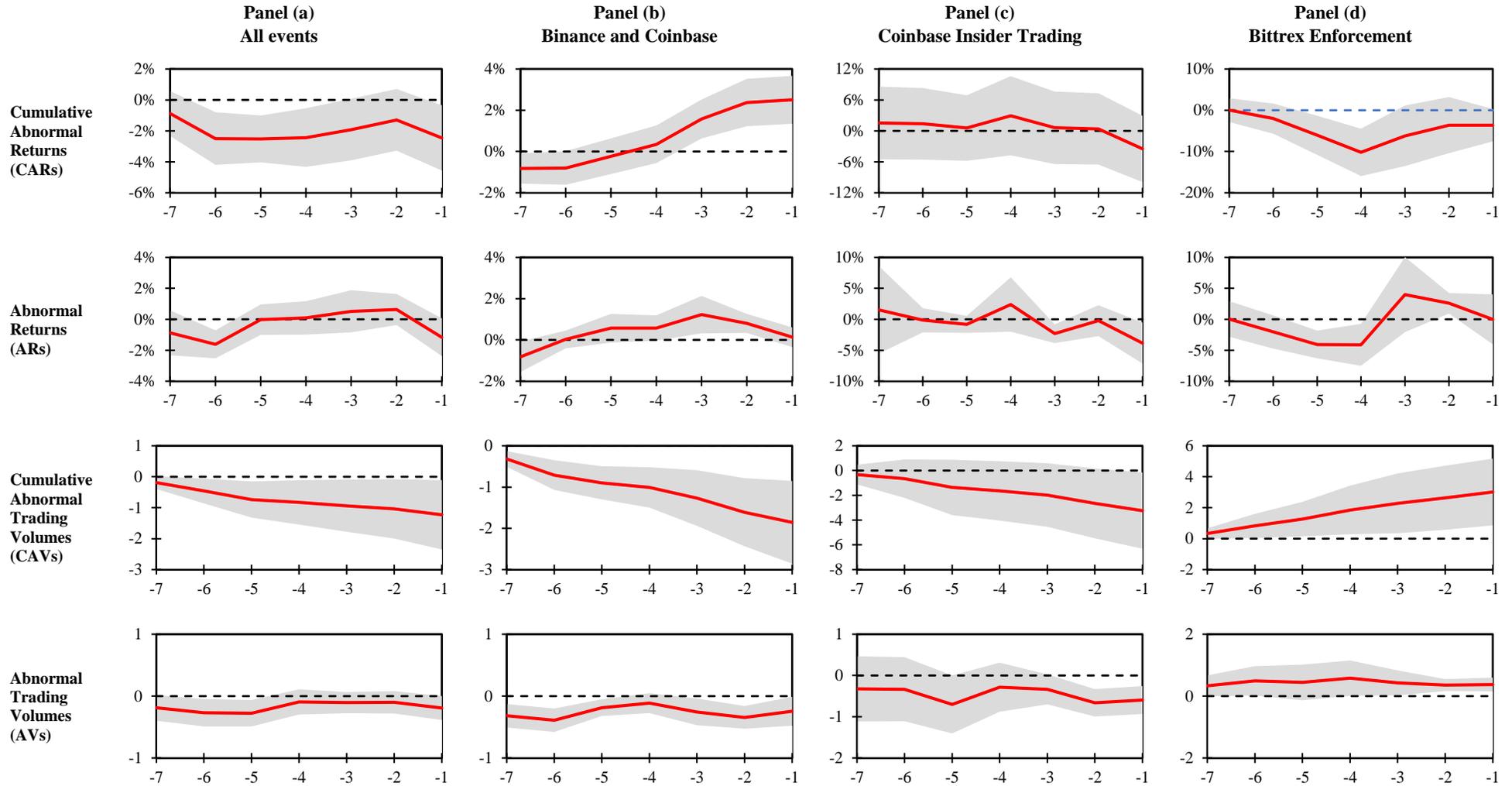